\documentclass[aps,showkeys,showpacs,notitlepage,superscriptaddress]{revtex4-1}
\pdfoutput=1
\usepackage{amsmath}
\usepackage{amssymb}
\usepackage{amsfonts}
\usepackage{graphicx}
\usepackage{hyperref}
\usepackage{color}
\usepackage{color}
\hypersetup{pdftitle=Forman curvature for directed networks}

\usepackage{mathtools}

\begin{document}
%-----------------------------------------------------------------
% Title
\title{Forman curvature for directed networks}
%-----------------------------------------------------------------
% Authors
\author{R.P. Sreejith}
\affiliation{The Institute of Mathematical Sciences, Homi Bhabha National Institute, Chennai, India}
\author{J\"urgen Jost}
\email{jost@mis.mpg.de}
\affiliation{Max Planck Institute for Mathematics in the Sciences, Leipzig, Germany}
\affiliation{The Santa Fe Institute, Santa Fe, New Mexico, USA}
\author{Emil Saucan}
\email{emil.saucan@mis.mpg.de}
\affiliation{Max Planck Institute for Mathematics in the Sciences, Leipzig, Germany}
\affiliation{Department of Electrical Engineering, Technion, Israel Institute of Technology, Haifa, Israel}
\author{Areejit Samal}
\email{asamal@imsc.res.in}
\affiliation{The Institute of Mathematical Sciences, Homi Bhabha National Institute, Chennai, India}
%-----------------------------------------------------------------

%-----------------------------------------------------------------
% Abstract
\begin{abstract}
A goal in network science is the geometrical characterization of complex networks. In this direction, we have recently introduced the  Forman's discretization of Ricci curvature to the realm of undirected networks. Investigation of this edge-centric network measure, Forman curvature, in diverse model and real-world undirected networks revealed that the curvature measure captures several aspects of the organization of complex undirected networks. However, many important real-world networks are inherently directed in nature, and the definition of the Forman curvature for undirected networks is unsuitable for the analysis of such directed networks. Hence, we here extend the Forman curvature for undirected networks to the case of directed networks. The simple mathematical formula for the Forman curvature of a directed edge elegantly incorporates node weights, edge weights and edge direction. By applying the Forman curvature for directed networks to a variety of model and real-world directed networks, we show that the measure can be used to characterize the structure of complex directed networks. Furthermore, our results also hold in real directed networks which are weighted or spatial in nature. These results in combination with our previous results suggest that the Forman curvature can be readily employed to study the organization of both directed and undirected complex networks.
\end{abstract}
%-----------------------------------------------------------------

%-----------------------------------------------------------------
\maketitle
%-----------------------------------------------------------------

%-----------------------------------------------------------------
% Introduction
\section{Introduction}
\label{Introduction}

Complex networks \cite{Watts1998,Barabasi1999,Albert2002,Newman2010,Fortunato2010,Dorogovtsev2013} are present everywhere in nature and society. Metabolic networks \cite{Jeong2000} capture interactions among different metabolites and enzymes that are responsible for growth and maintenance of a living organism.  Ecological networks \cite{Sole2001} capture interactions among different species in the ecosystem. Online networks such as Facebook \cite{Ellison2007} and Twitter \cite{Mcauley2012} capture relationships between different individuals in the society. Transportation networks \cite{Subelj2011} capture the movement of traffic across various parts of the world. Network science \cite{Watts1998,Barabasi1999,Albert2002,Newman2010,Fortunato2010,Dorogovtsev2013} aims to characterize the structure of these ubiquitous complex networks. Towards this goal, a recent focus in network science has been the development of geometry based measures to characterize the structure of complex networks \cite{Eckmann2002,Ollivier2009,Lin2010,Lin2011,Bauer2012,Jost2014,Wu2015,Ni2015,Sandhu2015a,Sreejith2016,Bianconi2017}.

A central concept in geometry is curvature which quantifies the deviation of an object from being flat. In differential geometry, where the notion of curvature originated, there are several types of curvature \cite{Jost2011}. Among those, it seems that the concept of Ricci curvature is the most useful for the analysis of graphs or networks. In differential geometry, Ricci curvature first appears in the so called Jacobi equation, that represents a linearization of the equation for geodesics \cite{Jost2011} and, as such, governs both the growth of volumes and the dispersion rate of geodesics \cite{Lott2009,Jost2011}. Ricci curvature is also an essential ingredient in the so called Bochner-Weitzenb\"{o}ck formula, that connects between the Laplacian on a manifold and its geometry, as expressed by its curvature. By its very definition through the above mentioned equation of geodesics, Ricci curvature depends on a direction, which then for networks translates into the fact that it should be a measure associated to an edge rather than to a node. Among the several curvature measures \cite{Eckmann2002,Shavitt2004,Saucan2005,Ollivier2009,Ollivier2010,Lin2010,Lin2011,Narayan2011,Bauer2012,Jost2014,Ollivier2013,Wu2015,Ni2015,Sandhu2015a,Sreejith2016} that have been proposed for geometrical characterization of complex networks, two different discretizations of the Ricci curvature, Ollivier-Ricci curvature \cite{Ollivier2009,Ollivier2010,Ollivier2013} and Forman-Ricci curvature \cite{Forman2003} seem particularly appealing and useful for analyses of complex networks. We remark that Ollivier-Ricci curvature directly captures, in a discrete setting, the growth of volumes property of Ricci curvature. In contrast, Forman-Ricci curvature stems from a discretization of the Bochner-Weitzenb\"{o}ck formula. Forman-Ricci curvature encodes, in a discrete setting, the dispersion of geodesics property of Ricci curvature.

\begin{figure}
\includegraphics[width=.7\columnwidth]{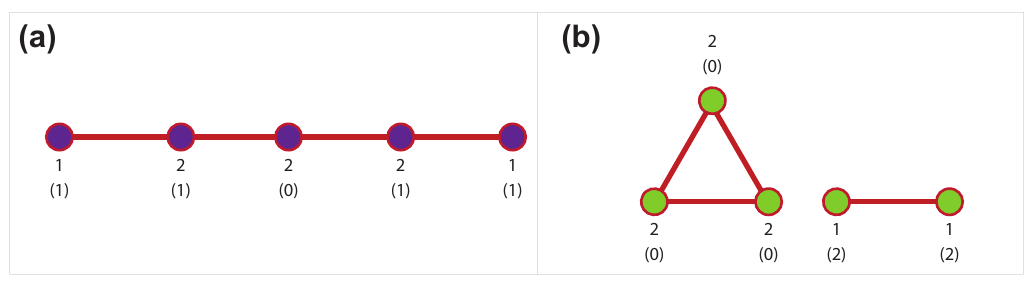}
\caption{Two undirected networks with the same degree distribution can have different distributions of Forman curvature. Each of the two networks shown in (a) and (b) have 5 nodes and 4 edges. The degree and Forman curvature is indicated besides each node with Forman curvature in parenthesis.}
\label{deg_cur}
\end{figure}

While Ollivier-Ricci curvature has already been systematically explored in complex undirected networks \cite{Lin2010,Lin2011,Bauer2012,Jost2014,Loisel2014,Ni2015,Sandhu2015a,Sandhu2015b}, we \cite{Sreejith2016} recently have introduced Forman's discretization of classical Ricci curvature \cite{Forman2003} to the realm of complex undirected networks. Astonishingly, the mathematical formula of the Forman curvature of an edge is enticingly simple which renders the measure suitable for analysis of large-scale networks \cite{Sreejith2016}. Importantly, the mathematical formula of the Forman curvature also elegantly incorporates both edge weights and node weights, and this also makes the measure suitable for analysis of both unweighted and weighted networks \cite{Sreejith2016}. Since Forman curvature represents a discretization of the classical Ricci curvature which is intrinsically associated with edges of a network, this notion of curvature does not necessitate the technical artifice of extending a measure for the curvature of nodes to the edges \cite{Sreejith2016}. Thus, Forman curvature can be exploited for edge-based analysis of complex networks. Given the definition of the Forman curvature of an edge, one can easily define the Forman curvature of a node in the network by summing or averaging the curvatures of its adjacent edges, somewhat analogous to the concept of scalar curvature in Riemannian geometry \cite{Jost2011}. We remark that the Forman curvature for an edge is a local measure dependent on weights of adjacent nodes and edges in the network \cite{Sreejith2016}. Still Forman curvature, a local geometric characteristic, can provide deep insights on the global topology of the network \cite{Forman2003}. Moreover, two networks with the same degree distribution can have very different distributions of Forman curvature (Fig.~\ref{deg_cur}).

Although, we have successfully introduced Forman curvature to undirected networks \cite{Sreejith2016}, several important real networks in nature and society are inherently directed in nature. These include the metabolic networks \cite{Jeong2000,Samal2006,Samal2011}, gene regulatory networks \cite{Milo2002}, signaling networks \cite{Maayan2005}, neural networks \cite{Sporns2004}, the world wide web (WWW) \cite{Broder2000}, online social networks \cite{Mcauley2012} and transportation networks \cite{Opsahl2010,Opsahl2011}. However, the two different discretizations of the Ricci curvature, Ollivier-Ricci curvature and Forman-Ricci curvature, have been developed and employed for the analysis of undirected graphs to date \cite{Lin2010,Lin2011,Bauer2012,Jost2014,Loisel2014,Ni2015,Sandhu2015a,Sandhu2015b,Sreejith2016}. To enable proper investigations of directed networks, we here extend the concept of Forman curvature to the domain of directed graphs. By investigating a variety of model and real directed networks, we show that our extension of the Forman curvature to directed graphs, which elegantly incorporates both edge weights and node weights, can be utilized to analyze and classify different types of directed networks. Thus, Forman curvature can hereafter be employed to investigate both undirected and directed complex networks.

The remainder of this paper is organized as follows. In the next section, we extend the notion of Forman curvature for undirected networks to directed networks. In subsequent section, we list the dataset of directed networks, both model and real, employed to investigate the Forman curvature for directed networks. In penultimate section, we describe our main results, and in the last section, we conclude with a summary and future outlook.

%-----------------------------------------------------------------

%-----------------------------------------------------------------
% Forman Curvature
\section{Mathematical definition}
\label{FormanCurvature}

In this section, we extend the notion of Forman curvature for undirected graphs to the setting of directed graphs, and present the simple mathematical formulas that allow the computation of the Forman curvature in directed networks.

Recently, we have introduced the Forman curvature to the realm of undirected networks, and showed that this simple yet powerful notion of curvature has several advantages over other curvature measures for networks \cite{Sreejith2016}. In our earlier contribution \cite{Sreejith2016} on undirected networks, we have elaborated on the geometric motivation and physical interpretation of the Forman curvature for networks. We remark that the classical Ricci curvature operates directionally along vectors, and in the discrete setting of networks, Forman curvature is associated with the discrete analogue of vectors, namely, edges. Forman curvature for an edge $e$ in the undirected graph is given by the following formula \cite{Sreejith2016}:
\begin{equation}
\label{FormanRicciEdge}
\mathbf{F}(e) = w_e \left( \frac{w_{v_1}}{w_e} +  \frac{w_{v_2}}{w_e}  - \sum_{e_{v_1}\ \sim\ e,\ e_{v_2}\ \sim\ e} \left[\frac{w_{v_1}}{\sqrt{w_e w_{e_{v_1} }}} + \frac{w_{v_2}}{\sqrt{w_e w_{e_{v_2} }}} \right] \right)\,
\end{equation}
In Eq. \ref{FormanRicciEdge}, $e$ denotes the edge under consideration between two nodes $v_1$ and $v_2$, $w_e$ denotes the weight of the edge $e$ under consideration, and $w_{v_1}$ and $w_{v_2}$ denote the weights associated with the nodes $v_1$ and $v_2$, respectively. $e_{v_1} \sim e$ and $e_{v_2} \sim e$ in Eq. \ref{FormanRicciEdge} denote the set of edges incident on nodes $v_1$ and $v_2$, respectively, after \textit{excluding} the edge $e$ under consideration which connects the two nodes $v_1$ and $v_2$. Note that the indices, $e_{v_1} \sim e$ and $e_{v_2} \sim e$, under the summation symbol in Eq. \ref{FormanRicciEdge} do not denote a double summation, but rather this notation enables compact representation of a single summation. That is,
\begin{equation}
\label{FormanRicciSingleSum}	
\sum_{e_{v_1}\ \sim\ e,\ e_{v_2}\ \sim\ e} \left[\frac{w_{v_1}}{\sqrt{w_e w_{e_{v_1} }}} + \frac{w_{v_2}}{\sqrt{w_e w_{e_{v_2} }}} \right]= \sum_{e_{v_1}\ \sim\ e} \frac{w_{v_1}}{\sqrt{w_e w_{e_{v_1}}}} + \sum_{e_{v_2}\ \sim\ e} \frac{w_{v_2}}{\sqrt{w_e w_{e_{v_2} }}}
\end{equation}

Using the definition of Forman curvature for edges in the undirected network, one can elegantly define the Forman curvature for a node $v$ in the undirected network as follows:
\begin{equation}
\label{FormanRicciNode}
\mathbf{F}(v) = \sum_{e_v\ \sim\ v} \mathbf{F}(e_v) \,
\end{equation}
where $e_v$ denotes the set of edges incident on the node $v$. In our earlier contribution \cite{Sreejith2016}, we had used a slightly different definition of the Forman curvature for a node $v$ where the sum of the Forman curvatures for edges incident on node $v$ in the undirected network was normalized by dividing the sum by degree of node $v$. Subsequently, we \cite{Sreejith2017} have found that the unnormalized definition of the Forman curvature for a node $v$ (Eq. \ref{FormanRicciNode}) has even better correlation with common network measures such as degree and betweenness centrality in undirected networks. Thus, we decided to use here the unnormalized definition of the Forman curvature for a node in the network.

\begin{figure}[!]
\includegraphics[width=.5\columnwidth]{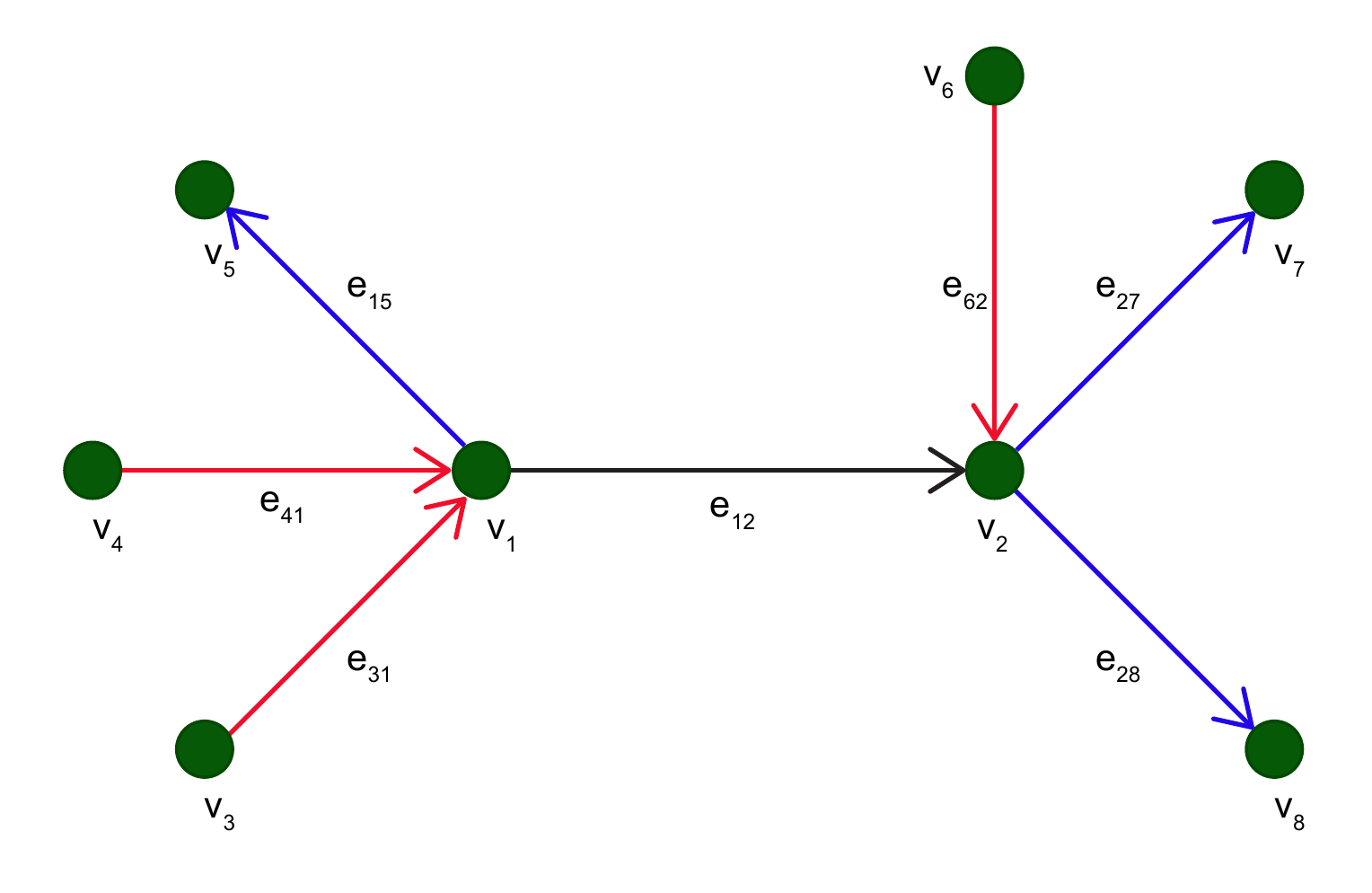}
\caption{Computation of the Forman curvature of a directed edge. In this example network, while computing the Forman curvature of the directed edge $e_{12}$ which originates at node $v_1$ and terminates at node $v_2$, we consider only the incoming edges $e_{31}$ and $e_{31}$ at node $v_1$ and only the outgoing edges $e_{27}$ and $e_{28}$ at node $v_2$ after excluding the edge $e_{12}$ under consideration. The \textit{incoming} and \textit{outgoing} edges for a node after excluding the edge under consideration are shown in red and blue, respectively, in the example network.}
\label{directed}
\end{figure}

In the original work \cite{Forman2003}, Forman had not envisaged his curvature function for directed networks, but rather concentrated on more general geometric objects, the so called CW weighted cell complexes, of which polygonal meshes, for instance, represent one well known and widely employed example. Notice that in Forman's original work only positive weights were considered, since these weights represent therein generalizations of such natural geometric notions as length, area and volume. In consequence, no directed surfaces (or, more general complexes) can be modeled directly using Forman's original formalism. However, we will show here that it is simple to adapt the Forman curvature for undirected graphs \cite{Sreejith2016} to directed graphs. In this aspect, our work, though restricted solely to networks, also represents a novel theoretical extension. In fact, when we rearrange the terms in the definition of the Forman curvature for an edge $e$ (Eq. \ref{FormanRicciEdge}), to separate the contributions of the two involved nodes $v_1$ and $v_2$,
\begin{equation}
\label{FormanRicciEdgeAlternate}
\mathbf{F}(e) = w_e \left( \frac{w_{v_1}}{w_e}   - \sum_{e_{v_1}\ \sim\ e} \frac{w_{v_1}}{\sqrt{w_e w_{e_{v_1} }}} \right)  + w_e \left( \frac{w_{v_2}}{w_e}   - \sum_{\ e_{v_2}\ \sim\ e} \frac{w_{v_2}}{\sqrt{w_e w_{e_{v_2} }}}  \right)\,
\end{equation}
we see that one can naturally define the curvature of a directed edge by only using the term involving its initial node, or alternatively that for its terminal node. While computing the Forman curvature of a directed edge $e = \overrightarrow{v_1v_2}$ that originates from node $v_1$ and terminates at node $v_2$, we take into account only those directed edges that either terminate at node $v_1$ or originate at node $v_2$ (Fig.~\ref{directed}). Moreover, while computing the Forman-Ricci curvature of the directed edge $e = \overrightarrow{v_1v_2}$, we ignore self-edges or self-loops to nodes $v_1$ and $v_2$, and also, the edge $\overrightarrow{v_2v_1}$ opposite to the one under consideration (in case such edges do occur in the considered network). In short, we consider only those neighboring edges in the computation of the Forman curvature in directed networks whose direction is compatible with the direction of edge $e$, and also, ensure that the resulting path is consistent with the direction of edge $e$ under consideration.

We remark that the definition adopted here for the Forman curvature of a directed edge is, by no means, the only possible choice. Beyond the mathematical reason discussed above, it is the most natural definition for the modeling of diverse real networks, such as, those considered in the results section. However, for the case of real networks with abundant feedback, e.g. gene regulatory networks or neural networks, it may be more realistic to consider self-edges. Note that self-loops are commonly drawn as directed arcs but they in fact are undirected, given that there is no preferential, intrinsic direction from a node to itself.

When we start with a node in a directed network, we can distinguish between its incoming and outgoing edges. Given a node $v$, let us denote the set of \textit{incoming} and \textit{outgoing} edges for a node $v$ by $E_{I,v}$ and $E_{O,v}$, respectively (Fig.~\ref{directed}). Then, one can elegantly define the \textit{In Forman curvature} $\mathbf{F}_I(v)$ and the \textit{Out Forman curvature} $\mathbf{F}_O(v)$ as follows:
\begin{equation}
\label{FormanRicciNodeIn}
\mathbf{F}_I(v) = {\sum_{e \in E_{I,v}} \mathbf{F}(e)}\,;
\end{equation}
\begin{equation}
\label{FormanRicciNodeOut}
\mathbf{F}_O(v) = {\sum_{e \in E_{O,v}} \mathbf{F}(e)}\,;
\end{equation}
where the summations are taken over only the incoming and outgoing edges, respectively. Moreover, one can obtain the total amount of flow through a node $v$ as follows:
\begin{equation}
\label{FormanRicciNodeFlow}
\mathbf{F}_{I/O}(v) =  \mathbf{F}_I(v) - \mathbf{F}_O(v)\,.
\end{equation}

%-----------------------------------------------------------------

%-----------------------------------------------------------------
% Datasets
\section{Datasets}
\label{datasets}

\noindent In this work, we have analyzed the Forman curvature for directed graphs in both model and real networks. We have considered two generative models for directed networks:
\begin{itemize}
\item \textbf{Erd\"{o}s-R\'{e}nyi (ER) model} \cite{Erdos1961} is commonly used to generate random graphs. ER model generates an ensemble $G(n,p)$ of graphs where $n$ is the number of nodes and $p$ is the probability that each possible directed edge exists between any pair of nodes in the network.
\item \textbf{Scale-free model} \cite{Bollobas2003} generates directed graphs with power-law degree distribution for both in-degree and out-degree of nodes. This growing network model implements a preferential attachment scheme wherein new nodes are connected to existing nodes based on in-degree and out-degree distribution. Starting from an initial network, the graph expansion at each discrete time step occurs through addition of new nodes or new edges. The graph expansion is based on three model parameters: $\alpha$, $\beta$ and $\gamma$. The parameter $\alpha$ gives the probability of adding a new node $v$ with an edge from $v$ to an existing node $w$ where node $w$ is chosen based on in-degree distribution. The parameter $\gamma$ gives the probability of adding a new node $v$ with an edge to $v$ from an existing node $w$ where node $w$ is chosen based on out-degree distribution. The parameter $\beta$ gives the probability of adding a new edge from an existing node $v$ to another existing node $w$, where $v$ and $w$ are chosen independently according to out-degree and in-degree distribution, respectively.
\end{itemize}

\noindent Apart from the above mentioned model networks, we have also considered the following directed and unweighted real networks:
\begin{itemize}
\item \textbf{\textit{E. coli} TRN} \cite{Salgado2013,Kumar2015} gives the transcriptional regulatory network in the bacterium \textit{Escherichia coli}. In this network of 3072 nodes and 7853 edges, nodes corresponds to genes and directed edges represent control of target gene expression through transcription factors.
\item \textbf{Air traffic control} is a directed network which was reconstructed based on the Preferred Routes Database of the US Federal Aviation Administration National Flight Data Center (NFDC). In this network of 1226 nodes and 2613 edges, nodes corresponds to airports or service centers and directed edges represent preferred routes recommended by the NFDC.
\item \textbf{Twitter Lists} \cite{Mcauley2012} is a directed network of connections between Twitter users. In this network of 23370 nodes and 33101 edges, nodes correspond to users and directed edges signify that the left user follows the right user on Twitter.
\item \textbf{Physicians} \cite{Coleman1957} is a directed network representing innovation among 246 physicians in different towns of Illinois, Peoria, Bloomington, Quincy and Galesburg in USA based on data collected in 1966. In this network of 241 nodes and 1098 edges, nodes corresponds to physicians and directed edges signify that the left physician mentioned that the right physician is his friend or that the left physician turns to the right physician for advice when needed.
\end{itemize}
Note that we take the weight of nodes and edges equal to 1 while computing the Forman curvature in above mentioned directed and unweighted networks.

\noindent In addition, we have also considered the following directed and weighted real networks:
\begin{itemize}
\item \textbf{US Airport} \cite{Opsahl2011} represents a directed network with positive edge weights which was generated based on flights between US airports in 2010. In this network of 1574 nodes and 28236 edges, nodes correspond to airports and directed edges represent flight connections between airports. The weight of each edge is proportional to the number of flights between the connecting cities.
\item \textbf{Advogato} \cite{Massa2009} is a directed network with positive edge weights which is based on an online community platform for developers of free software launched in 1999. In this network of 5155 nodes and 47135 edges, nodes correspond to users of Advogato and directed edges represent trust relationships between users. A trust link is called a certification on Advogato.
\item \textbf{Adolescent health} \cite{Moody2001} is a directed network with positive edge weights which was generated based on a survey in 1994-95 where each student was asked to list his or her 5 best female and 5 best male friends. In this network of 2539 nodes and 12969 edges, nodes correspond to students and directed edges between two students signify that the left student chose the right student as a friend.
\end{itemize}

\noindent Lastly, we have also considered the following directed and spatial real networks:
\begin{itemize}
\item \textbf{\textit{C. elegans} neural network} \cite{Kaiser2006} represents global neural network of the organism \textit{Caenorhabditis elegans}. In this network of 277 nodes and 2105 edges, nodes are neurons and directed edges give connections between the neurons. Note that this network is both directed and spatial unlike above mentioned real networks. In this network, edge weights correspond to the cartesian distance between the start and end nodes of an edge.
\item \textbf{Macaque neural network} \cite{Kaiser2006} gives the cortical connectivity network within one hemisphere of Macaque monkey. In this network of 94 nodes and 2390 edges, nodes correspond to cortical regions and directed edges signify links between each region. This network is also a spatial network, and edge weights correspond to the cartesian distance between the start and end nodes of an edge.
\end{itemize}

\noindent Several of the above mentioned real directed networks were downloaded from the KONECT \cite{Kunegis2013} database (Supplementary Table S1). The structure of the considered directed networks was analyzed using common network measures such as maximum degree, minimum degree, average degree, average in-degree, average out-degree, size of the largest weakly connected component and degree assortativity, and these results are contained in Supplementary Table S2.

%-----------------------------------------------------------------

%-----------------------------------------------------------------
% Results
\section{Results and Discussion}
\label{results}

%-----------------------------------------------------------------
% Curvature Distribution in Networks
\subsection{Curvature distribution in networks}
\label{distribution}

\begin{figure}[!]
\includegraphics[width=.7\columnwidth]{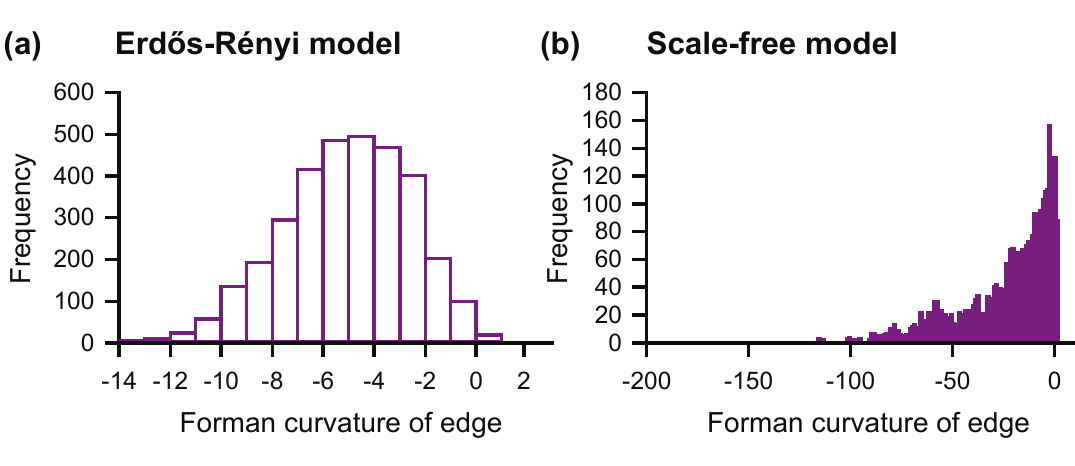}
\caption{Distribution of Forman curvature of directed edges in model networks. (a) Erd\"{o}s-R\'{e}nyi (ER) model with parameters: number of nodes $n=1000$ and probability $p$ that two nodes in the graph are directly connected $=0.00335$. (b) Scale-free model with parameters: number of nodes $n=1000$, probability of randomly adding a new node connected to an existing node based on the in-degree distribution $\alpha=0.142$, probability for adding an edge between two existing nodes $\beta=0.716$, and probability of randomly adding a new node connected to an existing node based on the out-degree distribution $\gamma=0.142$.}
\label{edge_dist_model}
\end{figure}

In Fig~\ref{edge_dist_model}, we show the distribution of the Forman curvature of directed edges in two models of directed networks. It is seen that most edges in both models of directed networks have negative curvature. One can clearly distinguish the two models of directed networks, ER and scale-free, by the observed nature of curvature distribution for edges (Fig.~\ref{edge_dist_model}). In random ER networks, the distribution of edge curvature is narrow with most edges having curvature close to -5. In scale-free networks, the distribution of edge curvature is broad with several edges having curvature less than -50. Also, random ER networks have narrow distribution of node curvatures while scale-free networks have broader distribution of node curvatures (Supplementary Figure S1). These results for Forman curvature in directed networks extend our recent results for undirected networks \cite{Sreejith2016}.

\begin{figure}[!]
\includegraphics[width=.7\columnwidth]{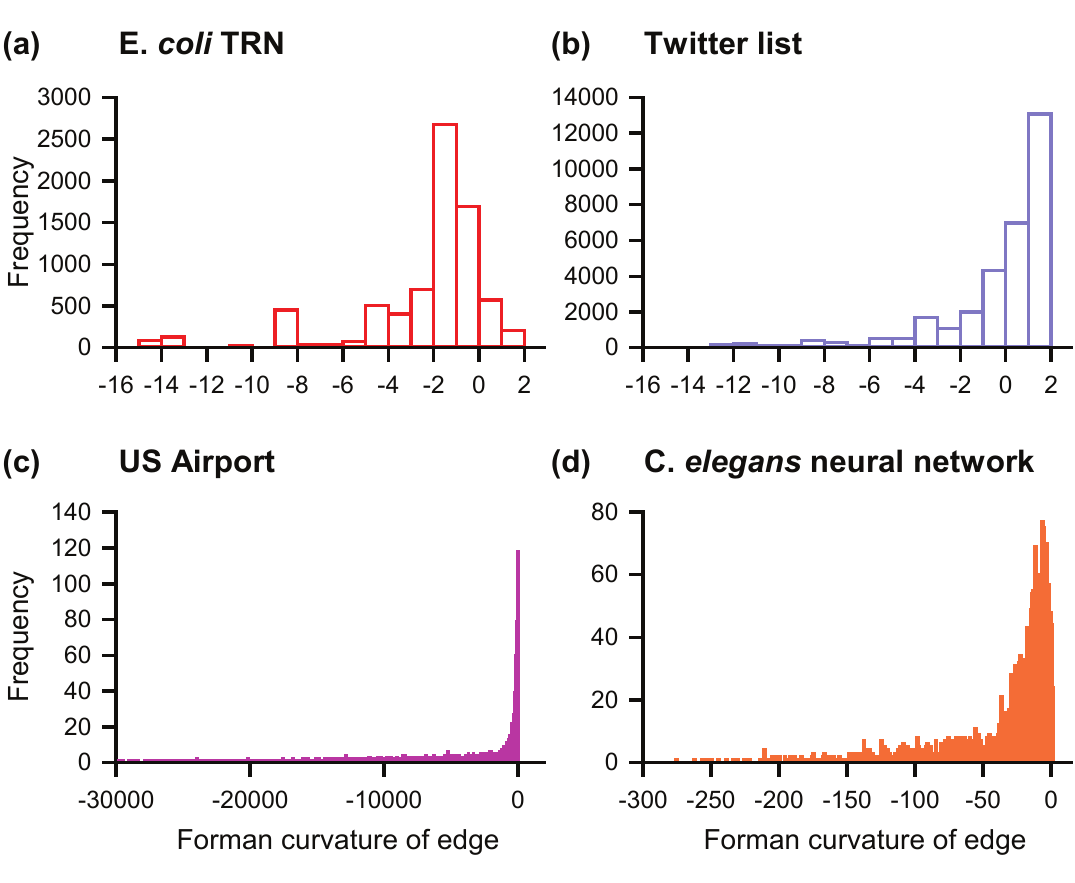}
\caption{Distribution of Forman curvature of directed edges in real networks. (a) \textit{E. coli} TRN. (b) Twitter Lists. (c) US Airport. (d) \textit{C. elegans} neural network.}
\label{edge_dist_real}
\end{figure}

In Fig~\ref{edge_dist_real}, we show the distribution of the Forman curvature of directed edges in real directed networks. Similar to models of directed networks, it is seen that most edges in considered real directed networks have negative curvature. The considered real directed networks have a broad distribution of edge curvature (Fig~\ref{edge_dist_real}). Also, it is clear that the nature of curvature distribution for edges in three considered real directed networks, Twitter Lists, US Airport and \textit{C. elegans} neural network, is similar to that of scale-free directed network. We remark that US Airports is a directed and weighted real network where edge weights are proportional to the number of flights between the connecting cities, while \textit{C. elegans} neural network is a directed and spatial network where edge weights correspond to the cartesian distance between the start and end nodes of an edge. Also, the distribution of node curvatures in three considered real directed networks, Twitter Lists, US Airport and \textit{C. elegans} neural network, is similar to that of scale-free directed network (Supplementary Figure S2). These observations are expected as most real networks have a scale-free architecture with power-law degree distribution \cite{Albert2002}.

Interestingly, the observed distribution of In Forman curvature is very different from that of Out Forman curvature in the \textit{E. coli} TRN (Supplementary Figure S2(a)). Bacterial transcriptional regulatory networks (TRNs) display an inherently hierarchical architecture \cite{Balazsi2005,Yu2006,Samal2008,Kumar2015} where few transcriptional factors at the top of the hierarchy have no incoming links (i.e., their in-degree is zero), while a large number of target genes at the bottom of the hierarchy have no outgoing links (i.e., their out-degree is zero). Moreover, the out-degree distribution of \textit{E. coli} TRN is broad with many global transcription factors regulating several target genes, while the in-degree distribution is narrow due to constraints on size of promoter regions in the bacterial genome \cite{Barabasi2004}. This asymmetry, but not the hierarchical structure by itself, seems to explain the observed difference in the distribution of In Forman curvature and Out Forman curvature in the \textit{E. coli} TRN (Supplementary Figure S2(a)).

Recall that Ricci curvature controls the growth of volumes in the classical Riemannian setting \cite{Berger2000,Jost2011}. Thus, spaces with negative Ricci curvature have exponential type growth while those with positive Ricci curvature have a finite diameter, and the result also holds for the Forman's discretization of the Ricci curvature \cite{Forman2003}. Thus, our observation that most edges and nodes in considered directed networks have a negative curvature suggests that these networks have potential of infinite growth.

%-----------------------------------------------------------------

%-----------------------------------------------------------------
% Edge curvature and other edge-based network measures
\subsection{Edge curvature and other edge-based network measures}

\begin{figure}[!]
\includegraphics[width=.7\columnwidth]{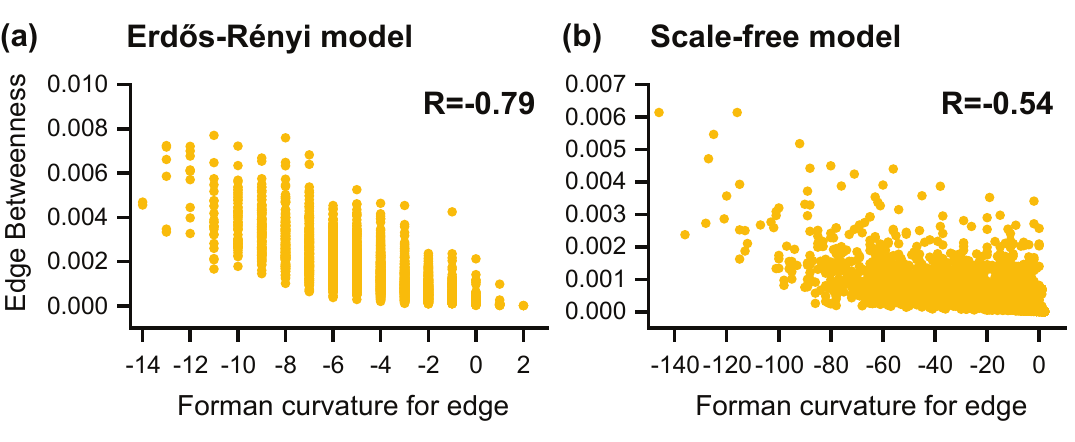}
\caption{Correlation between Forman curvature of directed edges and edge betweenness centrality in model networks. (a) Erd\"{o}s-R\'{e}nyi (ER) model. (b) Scale-free model. The parameters used to construct graphs from the generative models are same as those mentioned in Figure \ref{edge_dist_model}. We also indicate the Spearman correlation coefficient $R$ for each case.}
\label{cor_edge_model}
\end{figure}

\begin{figure}[!]
\includegraphics[width=.7\columnwidth]{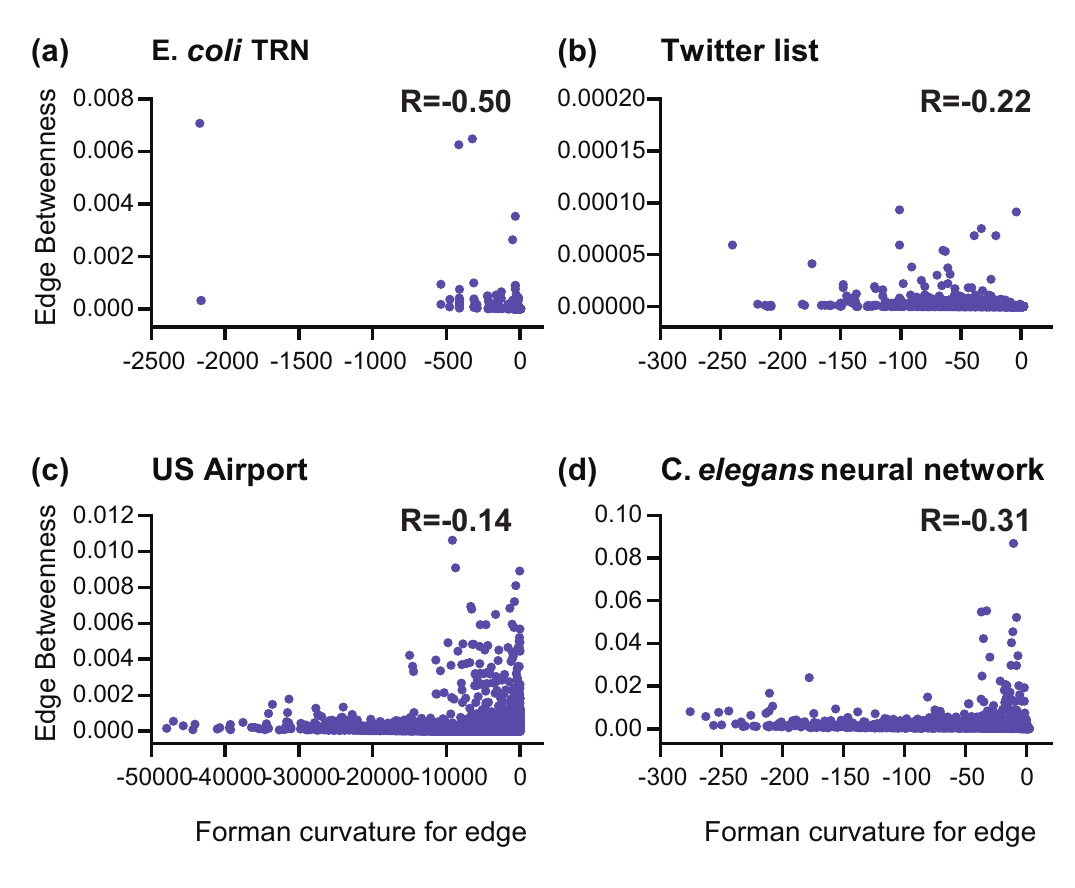}
\caption{Correlation between Forman curvature of directed edges and edge betweenness centrality in real networks. (a) \textit{E. coli} TRN. (b) Twitter Lists. (c) US Airport. (d) \textit{C. elegans} neural network. We also indicate the Spearman correlation coefficient $R$ for each case.}
\label{cor_edge_real}
\end{figure}

As emphasized above, Forman curvature is an edge-based measure for the analysis of complex networks \cite{Sreejith2016}. In network theory, edge-based measures are less common than node-based measures, and available edge-based measures for the analysis of undirected networks include edge betweenness \cite{Freeman1977,Brandes2001}, embeddedness \cite{Marsden1984} and dispersion \cite{Backstrom2014}. Among these edge-based measures for undirected networks, edge betweenness centrality can also be used to analyze directed networks. Edge betweennness centrality \cite{Freeman1977,Girvan2002,Newman2010} is a global measure that quantifies the number of shortest paths that pass through an edge in a network. Note that an edge with high betweenness centrality can be a bottleneck for flows in the network. We investigated the correlation between the two edge-based measures, Forman curvature and edge betweenness centrality, in model and real directed networks (Figs.~\ref{cor_edge_model} and \ref{cor_edge_real}; Supplementary Table S3). We find a high negative correlation between Forman curvature of an directed edge and edge betweenness centrality in the two models of directed networks (Fig.~\ref{cor_edge_model}; Supplementary Table S3). Moreover, in the considered real directed networks, moderate to high negative correlation is obtained between Forman curvature of a directed edge and edge betweenness centrality (Fig.~\ref{cor_edge_real}; Supplementary Table S3). Note that Forman curvature of an edge is a local measure dependent on weights of adjacent nodes and neighboring edges while edge betweenness centrality is a global measure dependent on all shortest paths in the network.

%-----------------------------------------------------------------

%-----------------------------------------------------------------
% Node curvature and common node-based network measures
\subsection{Node curvature and common node-based network measures}

In-degree of a node gives the number of edges incident to the node while out-degree of a node gives the number of edges originating from the node in a directed network. We show in Supplementary Figure S3 the correlation between in-degree and In Forman curvature of nodes, and in Supplementary Figure S4 the correlation between out-degree and Out Forman curvature of nodes in the two models of directed networks. We find a high negative correlation between In Forman curvature and in-degree of a node in the two models of directed networks (Supplementary Figure S3). A similar high negative correlation is obtained between Out Forman curvature and out-degree of a node in the two models of directed networks (Supplementary Figure S4). These results are expected because the In Forman curvature (respectively, the Out Forman curvature) of a node is computed based on the Forman curvature of directed edges incident on (respectively, originating from) the node, and the Forman curvature of a directed edge in the network is dependent on its neighboring edges. In particular, in the unweighted case, the curvature becomes more negative the higher the degree of the corresponding node. Also, these results for In Forman curvature and Out Forman curvature in the two models of directed networks reinforce our recent results for Forman curvature in models of undirected networks \cite{Sreejith2016}. Supplementary Figure S5 shows the correlation between in-degree and In Forman curvature of nodes, and Supplementary Figure S6 shows the correlation between out-degree and Out Forman curvature of nodes in real directed networks. Similar to models of directed networks, we find a high negative correlation between In Forman curvature and in-degree of a node (Supplementary Figure S5), and between Out Forman curvature and out-degree of a node (Supplementary Figure S6), in considered real directed networks. Supplementary Table S4 summarizes the results of these comparative analyses between In Forman curvature, Out Forman curvature, in-degree and out-degree of nodes in considered directed networks.

Betweenness centrality \cite{Freeman1977,Newman2010} of a node gives the fraction of shortest paths between all pairs of nodes in the network that pass through that node. Interestingly, high negative correlation is obtained between In Forman curvature and betweenness centrality of nodes (Supplementary Figure S3), and between Out Forman curvature and betweenness centrality of nodes (Supplementary Figure S4), in the two models of directed networks. Moreover, the magnitude of negative correlation between In Forman curvature and betweenness centrality of nodes is higher than the positive correlation between in-degree and betweenness centrality of nodes in the two models of directed networks. Also, the magnitude of negative correlation between Out Forman curvature and betweenness centrality of nodes is higher than the positive correlation between out-degree and betweenness centrality of nodes in the two models of directed networks. In the considered real directed networks, moderate to high negative correlation is obtained between In Forman curvature and betweenness centrality of nodes (Supplementary Figure S5), and between Out Forman curvature and betweenness centrality of nodes (Supplementary Figure S6). Similar to models of directed networks, the magnitude of negative correlation between In Forman curvature and betweenness centrality of nodes is higher than the positive correlation between in-degree and betweenness centrality of nodes in most of the considered real directed networks. Also, the magnitude of negative correlation between Out Forman curvature and betweenness centrality of nodes is higher than the positive correlation between out-degree and betweenness centrality of nodes in most of the considered real directed networks. Thus, we find that there is a high negative correlation between the global measure, betweenness centrality, and the local measures, In Forman curvature and Out Forman curvature, in model and real directed networks.

Pagerank \cite{Page1999,Langville2005} is an algorithm for directed networks which was originally developed to rank websites by the Google search engine. Pagerank can be employed with any directed network to measure importance of different nodes in the network. The algorithm counts the number and quality of incoming edges to a node to estimate the importance of a node in the network. Thus, we find as expected a high correlation between in-degree and pagerank of a node in the considered model and real directed networks. Interestingly, we find a high negative correlation between In Forman curvature and pagerank of a node in the considered model and real directed networks (Supplementary Table S4). We also find a moderate to high negative correlation between Out Forman curvature and pagerank in most of the considered model and real directed networks (Supplementary Table S4). Importantly, the magnitude of correlation between In Forman curvature and pagerank of a node in some of the weighted real directed networks is higher than the corresponding correlation between in-degree and pagerank of a node. These results emphasize that Forman curvature can be utilized to estimate the importance of nodes in complex networks. We remark that Forman curvature is naturally associated to edges. Still we have reported here the corresponding quantities for nodes, in order to make our results, better compatible with previous studies on network analysis.

%-----------------------------------------------------------------

%-----------------------------------------------------------------
% Robustness analysis
\subsection{Curvature and robustness of directed networks}
\label{robustness}

\begin{figure}[!]
\includegraphics[width=.7\columnwidth]{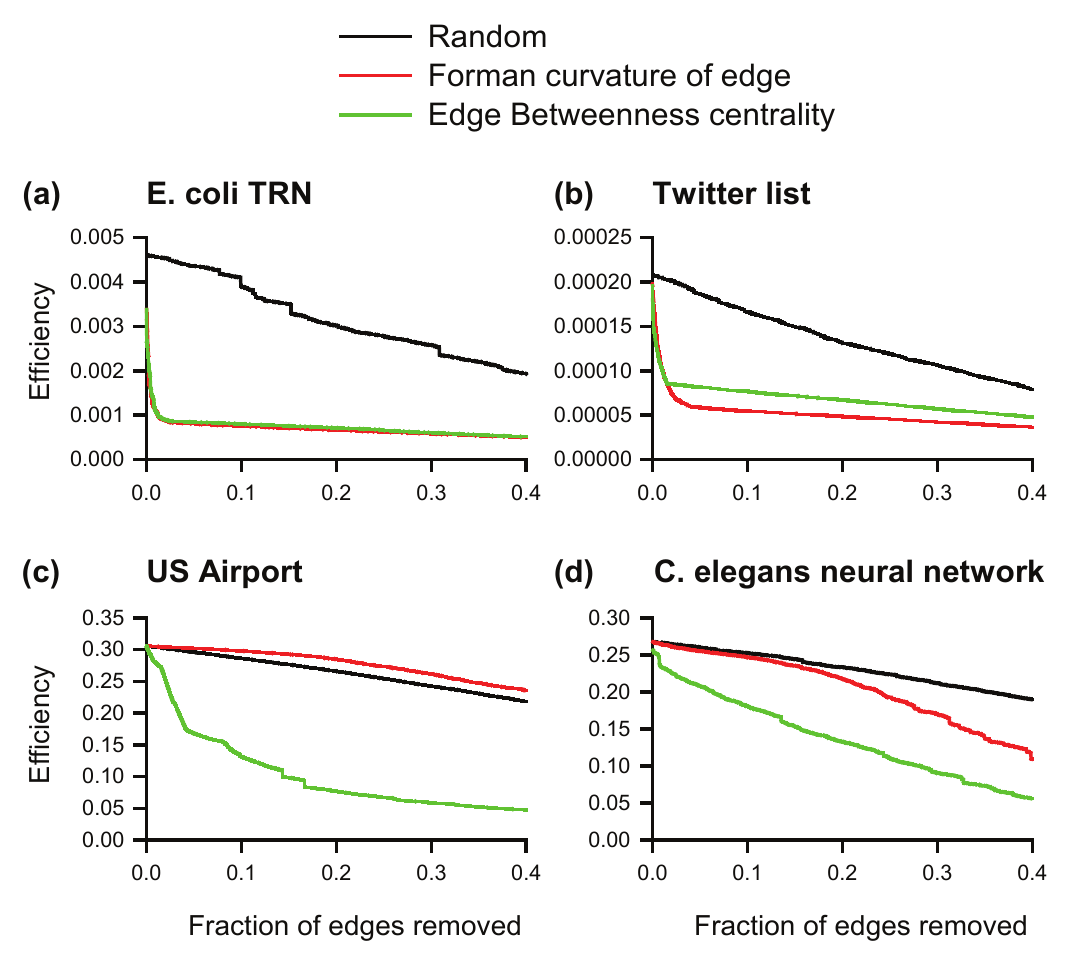}
\caption{Communication efficiency as a function of the fraction of edges removed in models of directed networks. (a) Erd\"{o}s-R\'{e}nyi (ER) model. (b) Scale-free model. In this figure, the order in which the edges are removed is based on the following criteria: Random order, Increasing order of Forman curvature, and Decreasing order of edge betweenness centrality. The parameters used to construct graphs from these generative models are same as those mentioned in Figure \ref{edge_dist_model}.}
\label{edge_rob_model}
\end{figure}

\begin{figure}[!]
\includegraphics[width=.7\columnwidth]{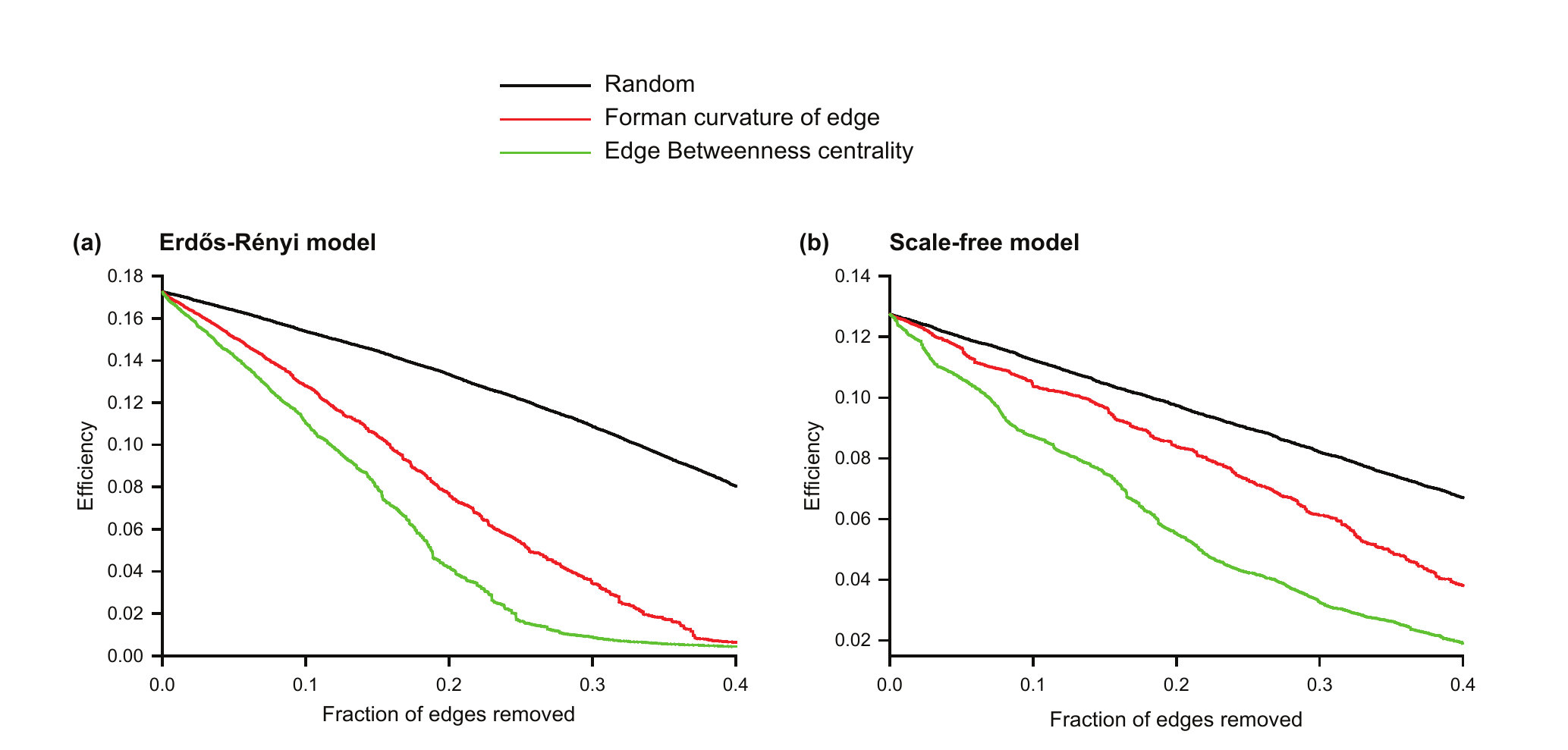}
\caption{Communication efficiency as a function of the fraction of edges removed in real directed networks. (a) \textit{E. coli} TRN. (b) Twitter Lists. (c) US Airport. (d) \textit{C. elegans} neural network.  In this figure, the order in which the edges are removed is based on the following criteria: Random order, Increasing order of Forman curvature, and Decreasing order of edge betweenness centrality.}
\label{edge_rob_real}
\end{figure}

We next investigated the effect of removing directed edges based on increasing order of their Forman curvature on the large-scale connectivity of directed networks. Communication efficiency \cite{Latora2001} is a measure that captures how efficiently the information can be exchanged across the network, and the measure can be used to quantify a network's resistance to failure in face of small perturbations. Fig.~\ref{edge_rob_model} shows the communication efficiency in two models of directed networks as a function of the fraction of edges removed. Here, the order of removing edges is based on the following criteria: (a) Random order, (b) Increasing order of Forman curvature (i.e, starting from the edge with most negative Forman curvature), and (c) Decreasing order of edge betweenness centrality. It is clear that targeted removal of edges with highly negative Forman curvature leads to faster disintegration compared to random removal of edges in the two models of directed networks (Fig.~\ref{edge_rob_model}). Similar to models of directed networks, we find that targeted removal of edges with highly negative Forman curvature leads to faster disintegration compared to random removal of edges in real directed networks (Fig.~\ref{edge_rob_real}). We have also compared the effect of removing edges based on increasing order of their Forman curvature against removing edges based on decreasing order of their edge betweenness centrality on the large-scale connectivity of networks (Figs.~\ref{edge_rob_model} and \ref{edge_rob_real}). In models of directed networks, we find that the removal of edges based on decreasing order of edge betweenness centrality leads to faster distintegration compared to removal of edges based on increasing order of Forman curvature (Fig.~\ref{edge_rob_model}). However, in some of the considered real directed networks, such as US Airport and \textit{C. elegans} neural network, the removal of edges based on decreasing order of edge betweenness centrality does not lead to faster distintegration compared to removal of edges based on increasing order of Forman curvature (Fig.~\ref{edge_rob_real}). We remark that US Airports and \textit{C. elegans} neural network are weighted in nature, and Forman curvature seems to perform better in weighted directed networks. Furthermore, the definition of Forman curvature naturally incorporates weights of both nodes and edges while the definition of edge betweenness centrality only incorporates the weights of edges. These results suggest that Forman curvature in future is likely to play a prominent role in the analysis of weighted networks.

We next investigated the effect of removing nodes based on increasing order of their In Forman curvature and Out Forman curvature on the large-scale connectivity of directed networks. It is seen that targeted removal of nodes with highly negative In Forman curvature or highly negative Out Forman curvature leads to faster disintegration compared to random removal of nodes in both model and real directed networks (Supplementary Figures S7 and S8). Previous work \cite{Barabasi1999,Jeong2000,Jeong2001,Albert2002,Joy2005,Yu2007,Newman2010} has shown that model and real networks are vulnerable to targeted removal of nodes with high degree or high betweenness centrality. We have also compared the effect of removing nodes based on increasing order of their In Forman curvature or Out Forman curvature against removing nodes based on decreasing order of their in-degree, out-degree or betweenness centrality on the large-scale connectivity of networks (Supplementary Figures S7 and S8). In models of directed networks, we find that the removal of nodes based on decreasing order of betweenness centrality leads to slightly faster distintegration compared to removal of nodes based on increasing order of In Forman curvature or Out Forman curvature (Supplementary Figure S7). In the considered real directed networks, we find that removal of nodes based on increasing order of In Forman curvature or Out Forman curvature leads to similar disintegration as observed with removal of nodes based on decreasing order of in-degree, out-degree or betweenness centrality (Supplementary Figure S8). Our results highlight that nodes with highly negative In Forman curvature or highly negative Out Forman curvature are important for maintaining the large-scale connectivity of directed networks.
%-----------------------------------------------------------------

%-----------------------------------------------------------------
% Node curvature and degree-degree correlations in directed networks
\subsection{Node curvature and degree-degree correlations in directed networks}
\label{deg_deg_correlations}

\begin{figure}[!]
\includegraphics[width=.7\columnwidth]{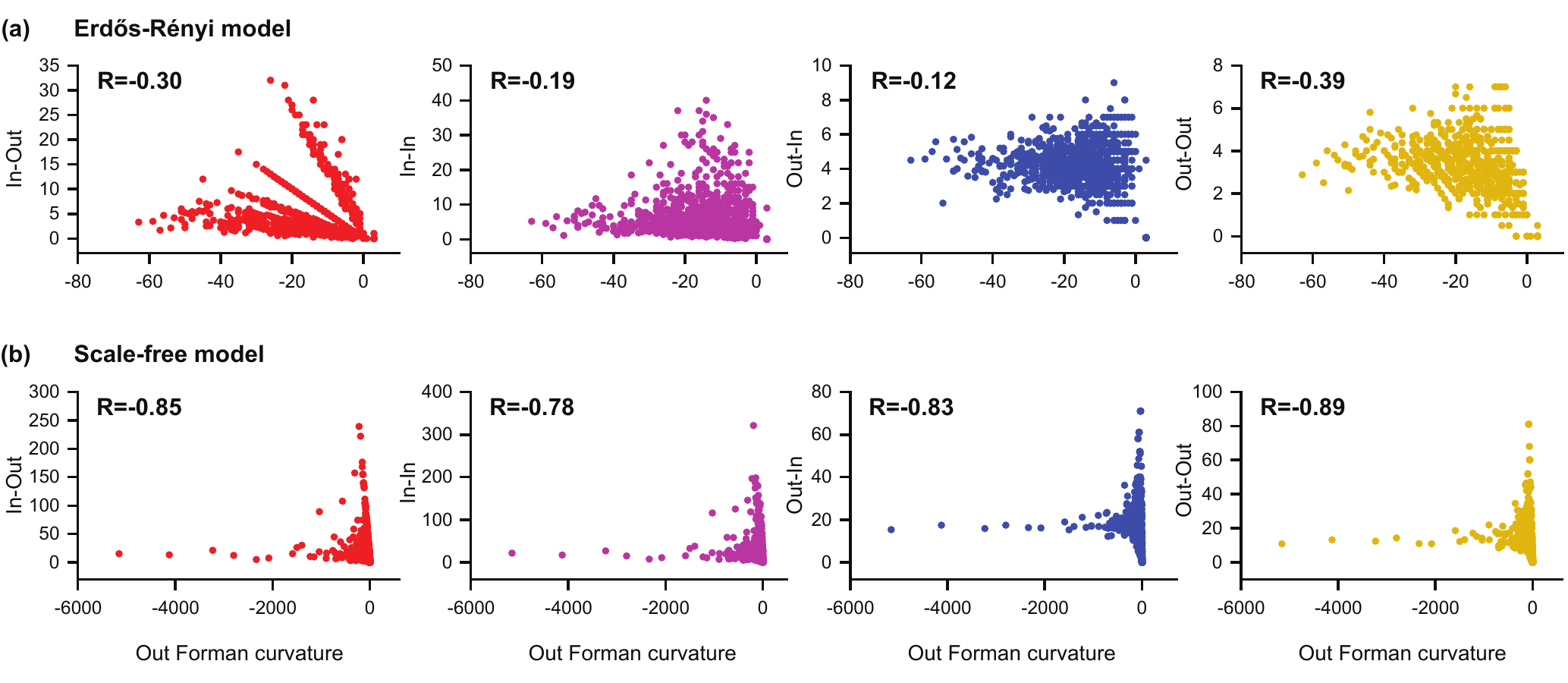}
\caption{Correlation between Out Forman curvature and the four assortativity coefficients in models of directed networks. (a) Erd\"{o}s-R\'{e}nyi (ER) model. (b) Scale-free model. The parameters used to construct graphs from the generative models are same as those mentioned in Figure \ref{edge_dist_model}. We also indicate the Spearman correlation coefficient $R$ for each case.}
\label{OF_deg_deg_model}
\end{figure}

Assortativity \cite{Newman2002} is a measure used to investigate degree-degree correlations in complex networks. A positive (negative) assortativity means that the degrees of neighboring nodes are positively (negatively) correlated. In a directed network, it is appropriate to distinguish the in-degree from out-degree of nodes, and this distinction naturally leads to four correlation (assortativity) coefficients corresponding to in-in, in-out, out-in and out-out degrees \cite{Foster2010}. We have investigated the correlation of the four assortativity coefficients with In Forman curvature and Out Forman curvature in both model and real directed networks. We find no correlation between the four assortativity coefficients and In Forman curvature in the two models of directed networks (Supplementary Table S4). In case of random ER networks, we find a small negative correlation between the four assortativity coefficients and Out Forman curvature, while in case of scale-free networks, we find a high negative correlation between the four assortativity coefficients and Out Forman curvature (Fig.~\ref{OF_deg_deg_model}; Supplementary Table S4). In case of considered real directed networks, we find either no correlation or moderate negative correlation between the four assortativity coefficients and In Forman curvature (Supplementary Table S4), while we find moderate to high negative correlation between the four assortativity coefficients and Out Forman curvature (Fig.~\ref{OF_deg_deg_real}). In Supplementary Table S5, we also report the correlation of the four assortativity coefficients with in-degree and out-degree in considered model and real directed networks.

%-----------------------------------------------------------------

%-----------------------------------------------------------------
% Conclusions
\section{Conclusions}
\label{conclusion}

In this paper, we have extended our recent adaptation of the Forman curvature for undirected networks to directed networks. The mathematical expression for the
Forman curvature of a directed edge elegantly incorporates the node weights, edge weights and edge direction in directed networks. The distribution of the Forman curvature of directed edges is narrow in random directed networks, while the distribution is broad in scale-free directed networks. In most real directed networks, the distribution of the Forman curvature of directed edges is also broad like scale-free networks. These results highlight that the distribution of the Forman curvature of directed edges can also be employed to distinguish and classify different types of directed networks.  We next investigated the correlation between Forman curvature of directed edges and edge betweenness centrality in considered directed networks. We find a significant negative correlation between Forman curvature of directed edges and edge betweenness centrality in considered directed networks. By investigating the effect of removing edges based on their Forman curvature on the communication efficiency of networks, we show that both model and real directed networks are vulnerable to targeted removal of edges with highly negative Forman curvature. Based on the definition of the Forman curvature of a directed edge, it is easy to define the In Forman curvature (Eq.~\ref{FormanRicciNodeIn}) and the Out Forman curvature (Eq.~\ref{FormanRicciNodeOut}) for nodes in directed networks. We also investigated the correlation of In Forman curvature and Out Forman curvature of nodes with common node-based measures such as in-degree, out-degree and betweenness centrality in considered directed networks. Finally, we investigated the correlation of the four assortativity coefficients in directed networks with In Forman curvature and Out Forman curvature in considered directed networks. Importantly, we have shown that the above results hold also for real directed networks which are weighted or spatial in nature. Also, the results reported here for directed networks mirror our recent results \cite{Sreejith2016} for undirected networks. Hence, Forman curvature can hereafter be employed to study the structure of both directed and undirected complex networks.

Recently, two different discretizations of the classical Ricci curvature, Ollivier-Ricci curvature \cite{Ollivier2009,Ollivier2010,Ollivier2013} and Forman-Ricci curvature \cite{Forman2003} have been introduced into graph theory. Whereas Ollivier's curvature has already been systematically investigated and also applied for the study of complex networks \cite{Lin2010,Lin2011,Bauer2012,Jost2014,Loisel2014,Ni2015,Sandhu2015a,Sandhu2015b}, in \cite{Sreejith2016}, we have initiated the investigation of the Forman curvature of empirical networks. While the computation of Forman's curvature in undirected networks is extremely simple, the computation of the more established Ollivier's curvature in undirected networks necessitates solving a linear programming problem associated with optimal mass transport on networks. Hence, the computation of Ollivier's curvature unlike Forman's curvature may not easily scale to extremely large networks such as the world wide web (WWW). Importantly, we have extended here the Forman curvature for undirected networks to the domain of directed networks in a very natural manner, while such an extension of Ollivier's curvature to directed networks is still awaited. Based on our results and the simplicity of the mathematical formulas associated with Forman curvature, we expect this novel curvature measure to be widely used for geometrical characterization of networks.

Previously, the clustering coefficient \cite{Watts1998} has been used as a reference measure to quantify curvature in complex undirected networks \cite{Bridson1999,Eckmann2002}. The clustering coefficient was originally designed for undirected graphs but the concept has been subsequently extended to directed graphs \cite{Fagiolo2007}. In our earlier analyses of undirected networks \cite{Sreejith2016}, we had found no or weak correlation between Forman curvature and clustering coefficient in the considered model and real undirected networks. Moreover, Forman curvature in undirected networks as opposed to clustering coefficient was found to have a significant correlation with degree and centrality measures \cite{Sreejith2016}. A natural  future direction will be the investigation of  the level of association between directed clustering coefficient \cite{Fagiolo2007} and In Forman curvature or Out Forman curvature in directed networks. In this direction, let us conclude with the following concrete suggestion. The definition of the Forman curvature is coupled with two fitting Laplacians. One should try to employ these Laplacians associated with the Forman curvature for the denoising of directed biological networks such as transcriptional regulatory networks and signalling networks.

%-----------------------------------------------------------------

%-----------------------------------------------------------------
% Acknowledgments
\section*{Acknowledgments}

ES and AS thank the Max Planck Institute for Mathematics in the Sciences, Leipzig, for their warm hospitality. AS would also like to thank Anand Pathak, M. Karthikeyan, R.P. Vivek-Ananth for discussions, and acknowledges support from Max Planck Society, IMSc PRISM project (XII Plan), Ramanujan fellowship (SB/S2/RJN-006/2014) and Department of Science and Technology (DST) start-up project (YSS/2015/000060).
%-----------------------------------------------------------------

%-----------------------------------------------------------------
% Bibliography

\pagebreak
%-----------------------------------------------------------------
% Figures
%-----------------------------------------------------------------
% Figure 1
%-----------------------------------------------------------------
%-----------------------------------------------------------------
% Figure 2
%-----------------------------------------------------------------
%-----------------------------------------------------------------
% Figure 3
%-----------------------------------------------------------------
%-----------------------------------------------------------------
% Figure 4
%-----------------------------------------------------------------
%-----------------------------------------------------------------
% Figure 5
%-----------------------------------------------------------------
%-----------------------------------------------------------------
% Figure 6
%-----------------------------------------------------------------
%-----------------------------------------------------------------
% Figure 7
%-----------------------------------------------------------------
%-----------------------------------------------------------------
% Figure 8
%-----------------------------------------------------------------
%-----------------------------------------------------------------
% Figure 9
%-----------------------------------------------------------------
%-----------------------------------------------------------------
% Figure 10
\begin{figure}[!]
\includegraphics[width=.7\columnwidth]{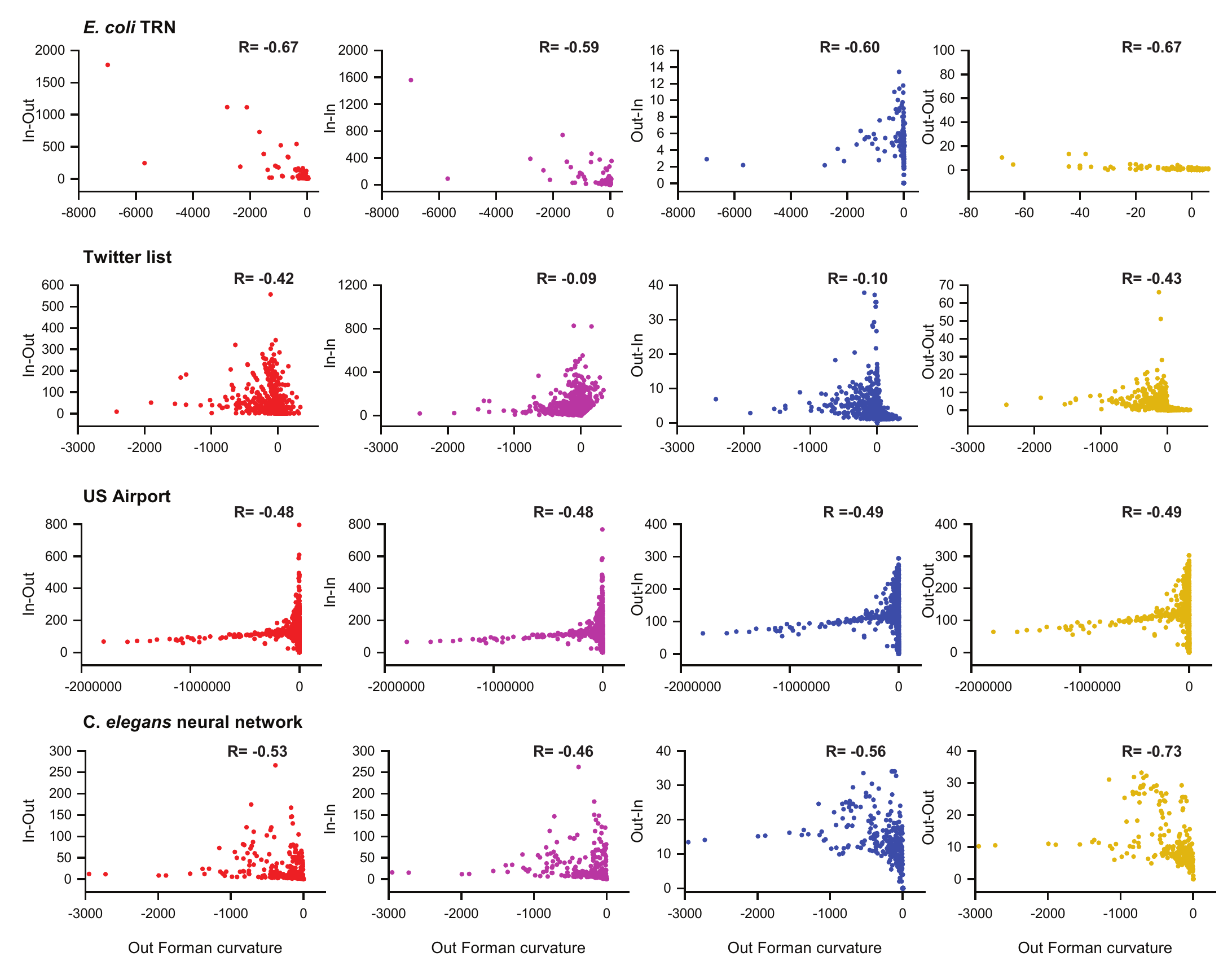}
\caption{Correlation between Out Forman curvature and the four assortativity coefficients in real directed networks. (a) \textit{E. coli} TRN. (b) Twitter Lists. (c) US Airport. (d) \textit{C. elegans} neural network.  We also indicate the Spearman correlation coefficient $R$ for each case.}
\label{OF_deg_deg_real}
\end{figure}
%-----------------------------------------------------------------

%-----------------------------------------------------------------
% Supplementary Material
\newpage
\setcounter{figure}{0}
\makeatletter
\renewcommand{\thefigure}{S\arabic{figure}}
%-----------------------------------------------------------------

%-----------------------------------------------------------------
% Supplementary Figures
\begin{figure}[!]
\includegraphics[width=.7\columnwidth]{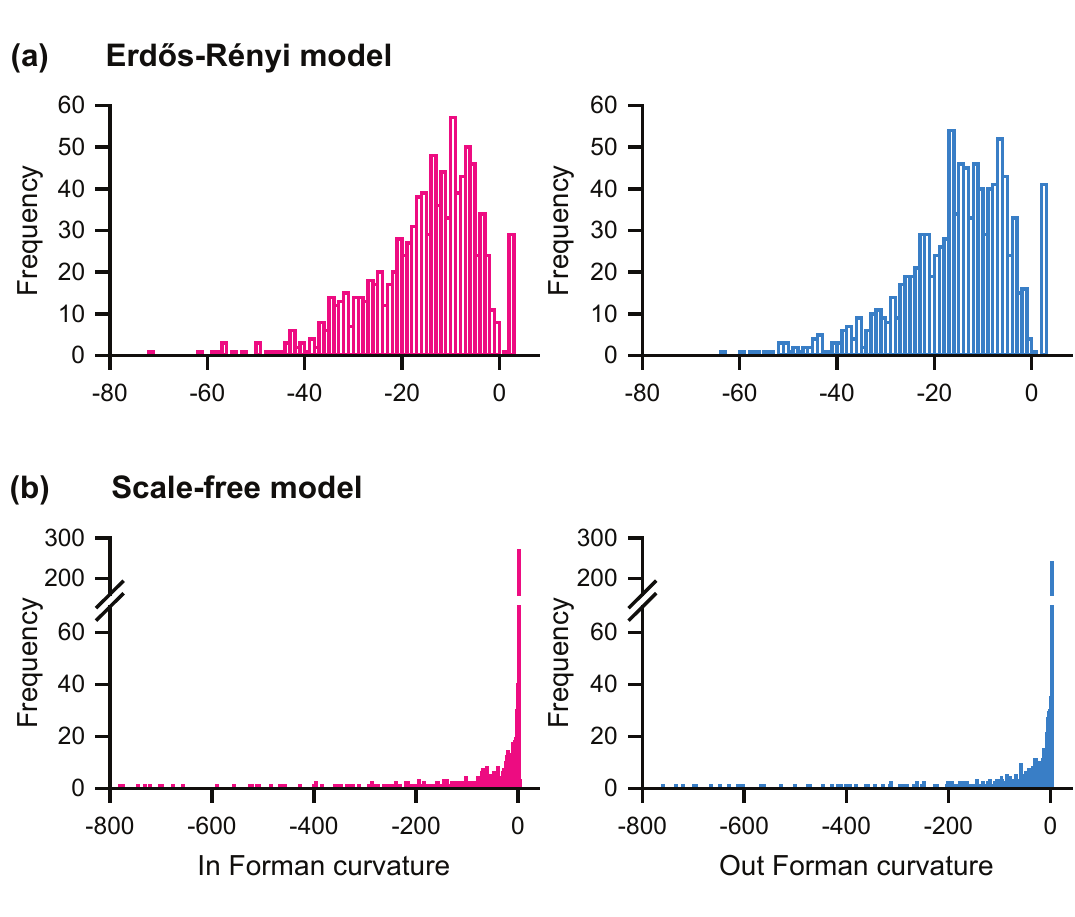}
\caption{Distribution of In Forman curvature and Out Forman curvature of nodes in model networks. (a) Erd\"{o}s-R\'{e}nyi (ER) model. (b) Scale-free model. The parameters used to construct graphs from the generative models are same as those mentioned in Figure 3.}
\end{figure}

\begin{figure}[!]
\includegraphics[width=.7\columnwidth]{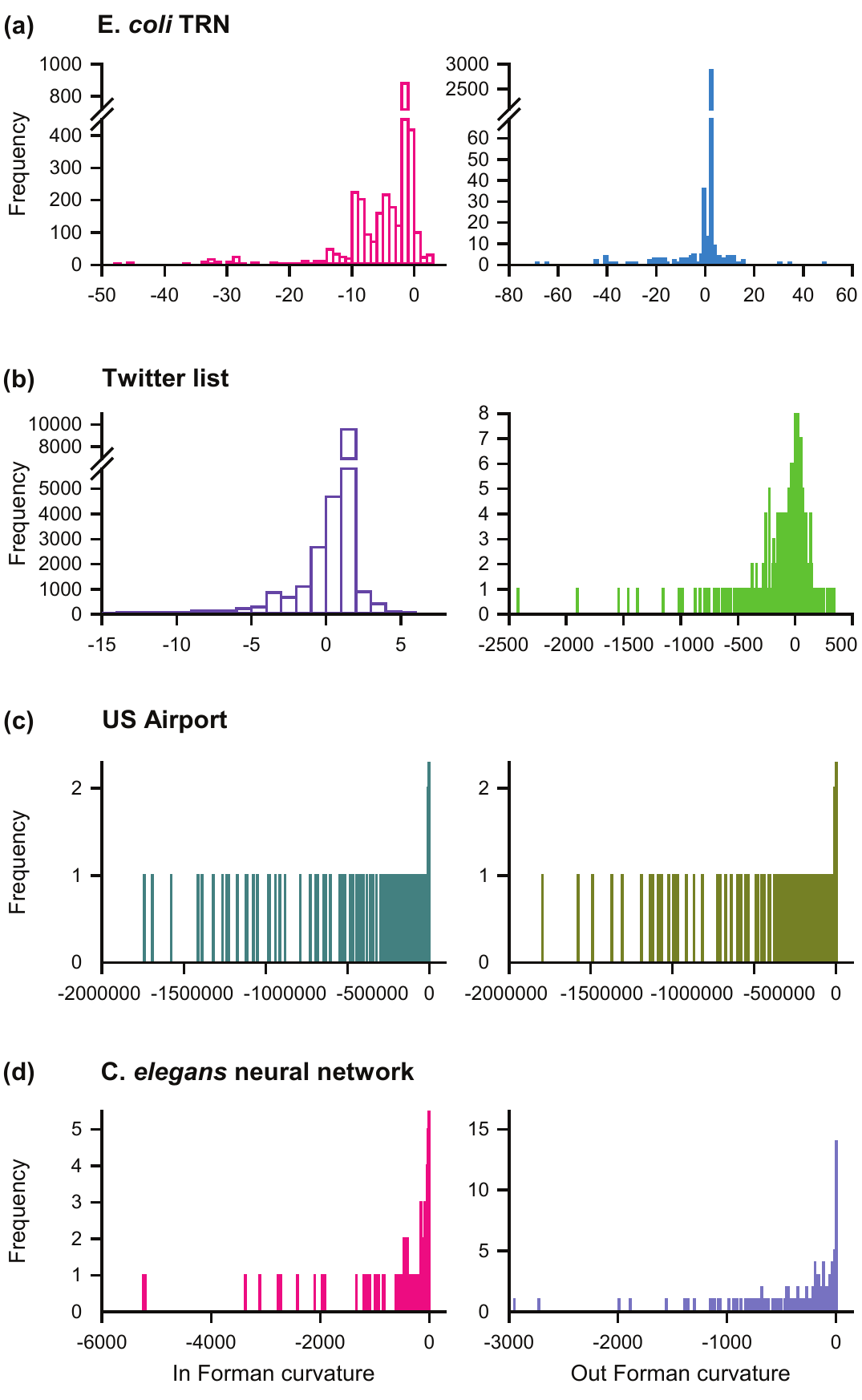}
\caption{Distribution of In Forman curvature and Out Forman curvature of nodes in real networks. (a) \textit{E. coli} TRN. (b) Twitter Lists. (c) US Airport. (d) \textit{C. elegans} neural network.}
\end{figure}

\begin{figure}[!]
\includegraphics[width=.7\columnwidth]{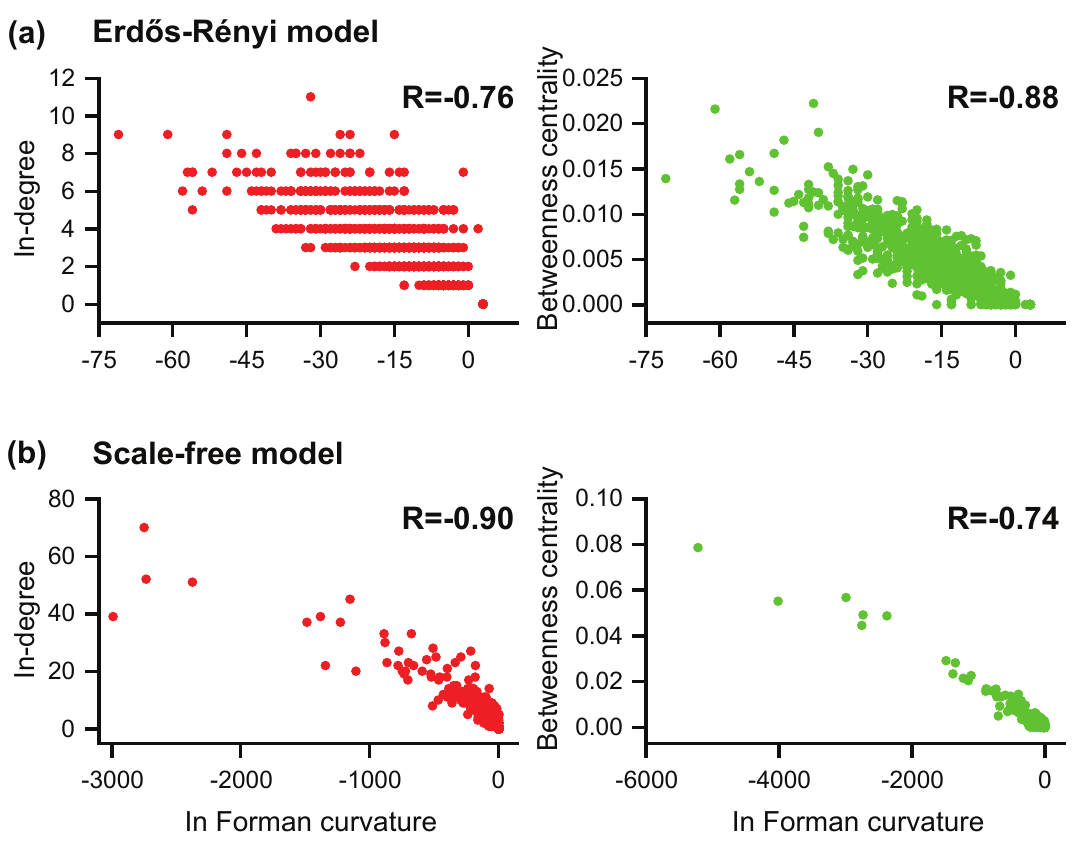}
\caption{Correlation between In Forman curvature and in-degree or betweenness centrality of nodes in model networks. (a) Erd\"{o}s-R\'{e}nyi (ER) model. (b) Scale-free model. The parameters used to construct graphs from the generative models are same as those mentioned in Figure 3. We also indicate the Spearman correlation coefficient $R$ for each case.}
\end{figure}

\begin{figure}[!]
\includegraphics[width=.7\columnwidth]{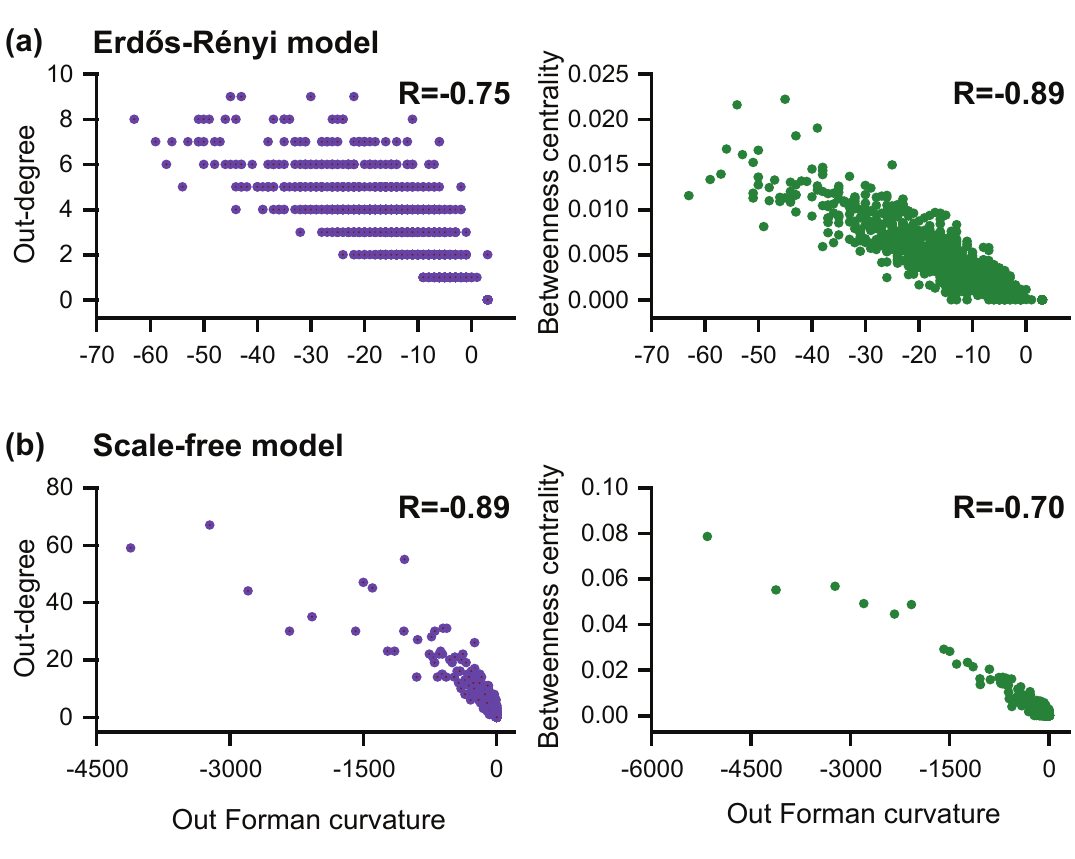}
\caption{Correlation between Out Forman curvature and out-degree or betweenness centrality of nodes in model networks. (a) Erd\"{o}s-R\'{e}nyi (ER) model. (b) Scale-free model. The parameters used to construct graphs from the generative models are same as those mentioned in Figure 3. We also indicate the Spearman correlation coefficient $R$ for each case.}
\end{figure}

\begin{figure}[!]
\includegraphics[width=.7\columnwidth]{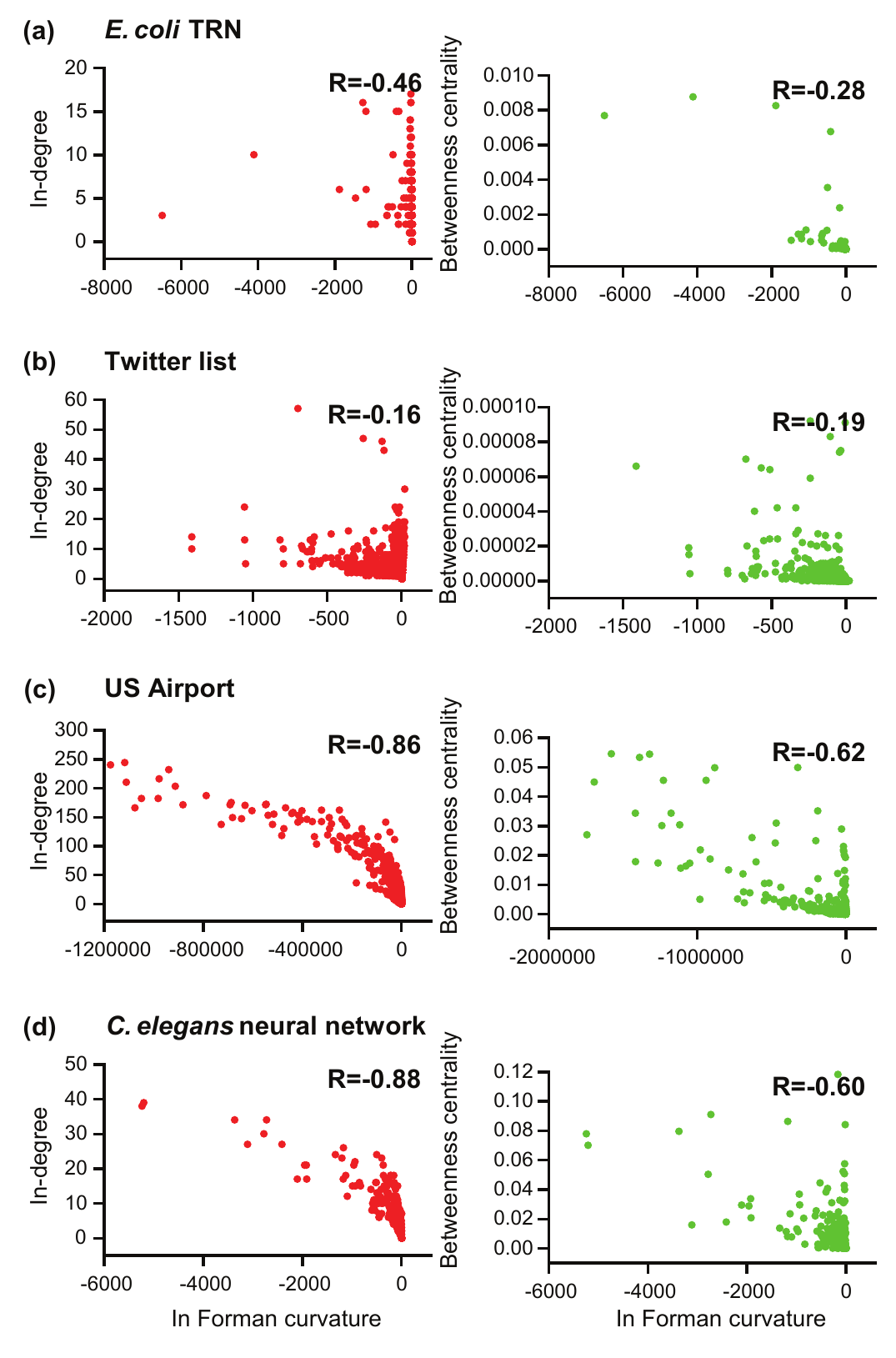}
\caption{Correlation between In Forman curvature and in-degree or betweenness centrality of nodes in real networks. (a) \textit{E. coli} TRN. (b) Twitter Lists. (c) US Airport. (d) \textit{C. elegans} neural network. We also indicate the Spearman correlation coefficient $R$ for each case.}
\end{figure}

\begin{figure}[!]
\includegraphics[width=.7\columnwidth]{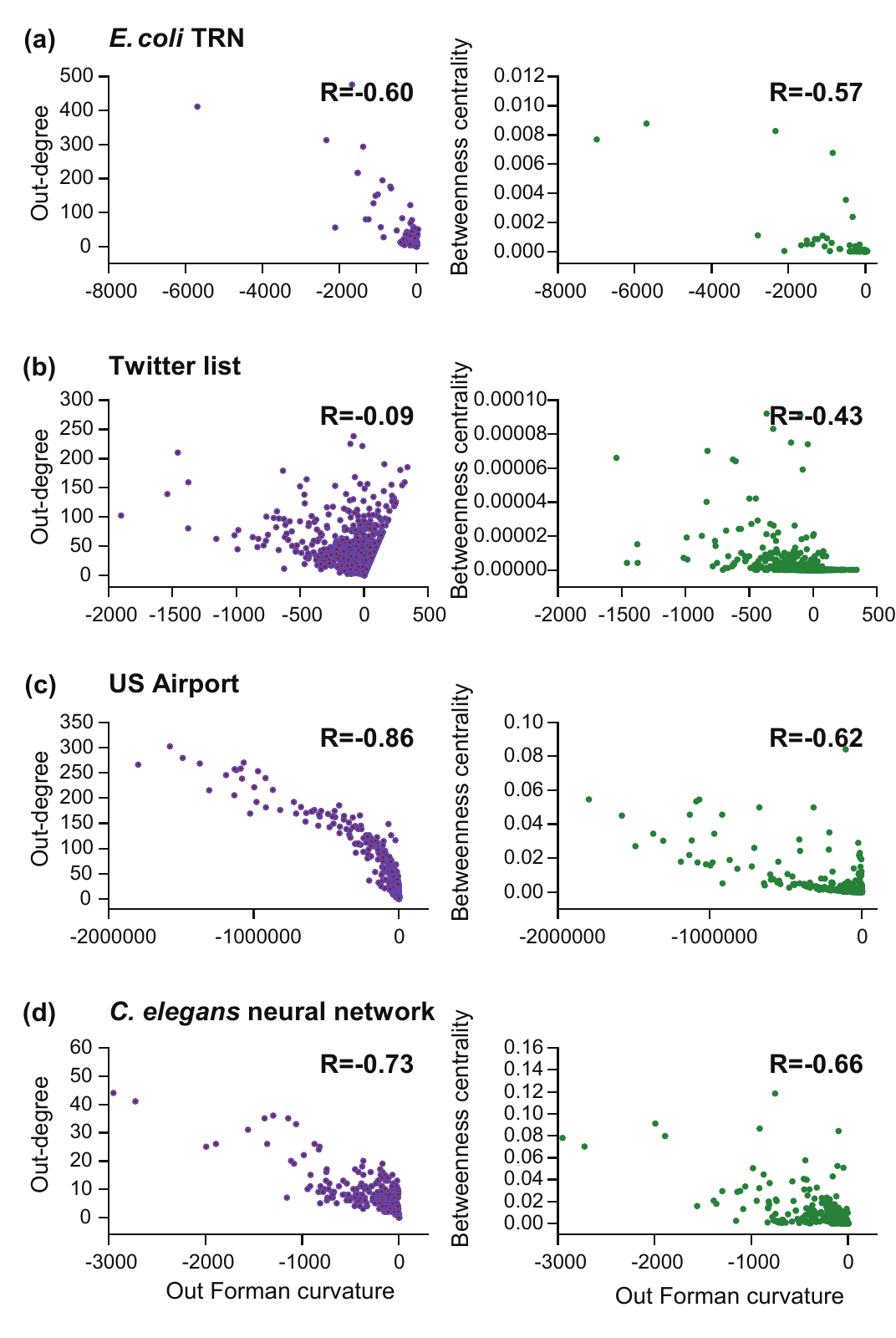}
\caption{Correlation between Out Forman curvature and out-degree or betweenness centrality of nodes in real networks. (a) \textit{E. coli} TRN. (b) Twitter Lists. (c) US Airport. (d) \textit{C. elegans} neural network. We also indicate the Spearman correlation coefficient $R$ for each case.}
\end{figure}

\begin{figure}[!]
\includegraphics[width=.7\columnwidth]{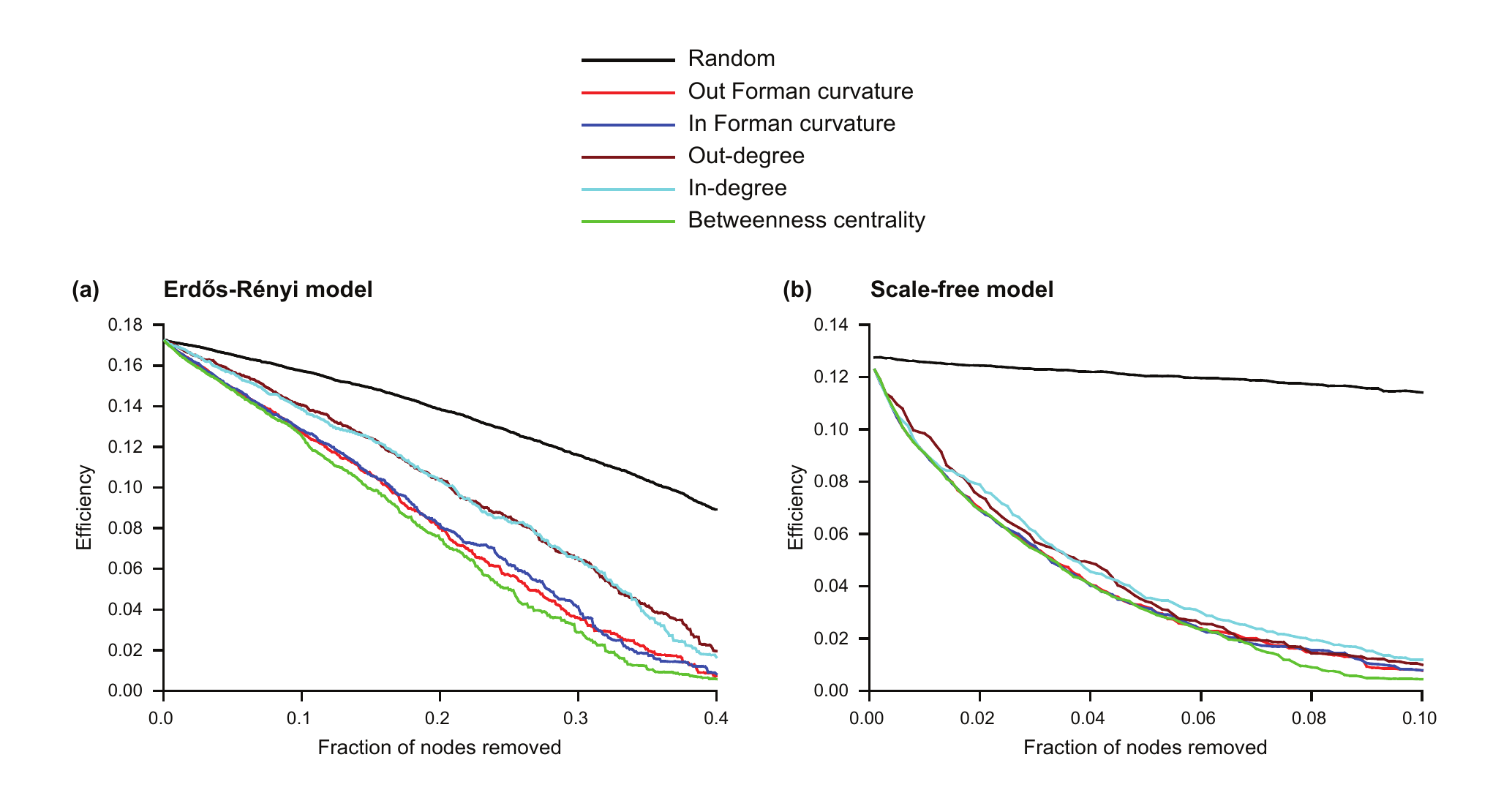}
\caption{Communication efficiency as a function of the fraction of nodes removed in models of directed networks. (a) Erd\"{o}s-R\'{e}nyi (ER) model. (b) Scale-free model. In this figure, the order in which the nodes are removed is based on the following criteria: Random order, Increasing order of In Forman curvature, Increasing order of Out Forman curvature, Decreasing order of in-degree, Decreasing order of out-degree, and Decreasing order of betweenness centrality. The parameters used to construct graphs from these generative models are same as those mentioned in Figure 3.}
\end{figure}

\begin{figure}[!]
\includegraphics[width=.7\columnwidth]{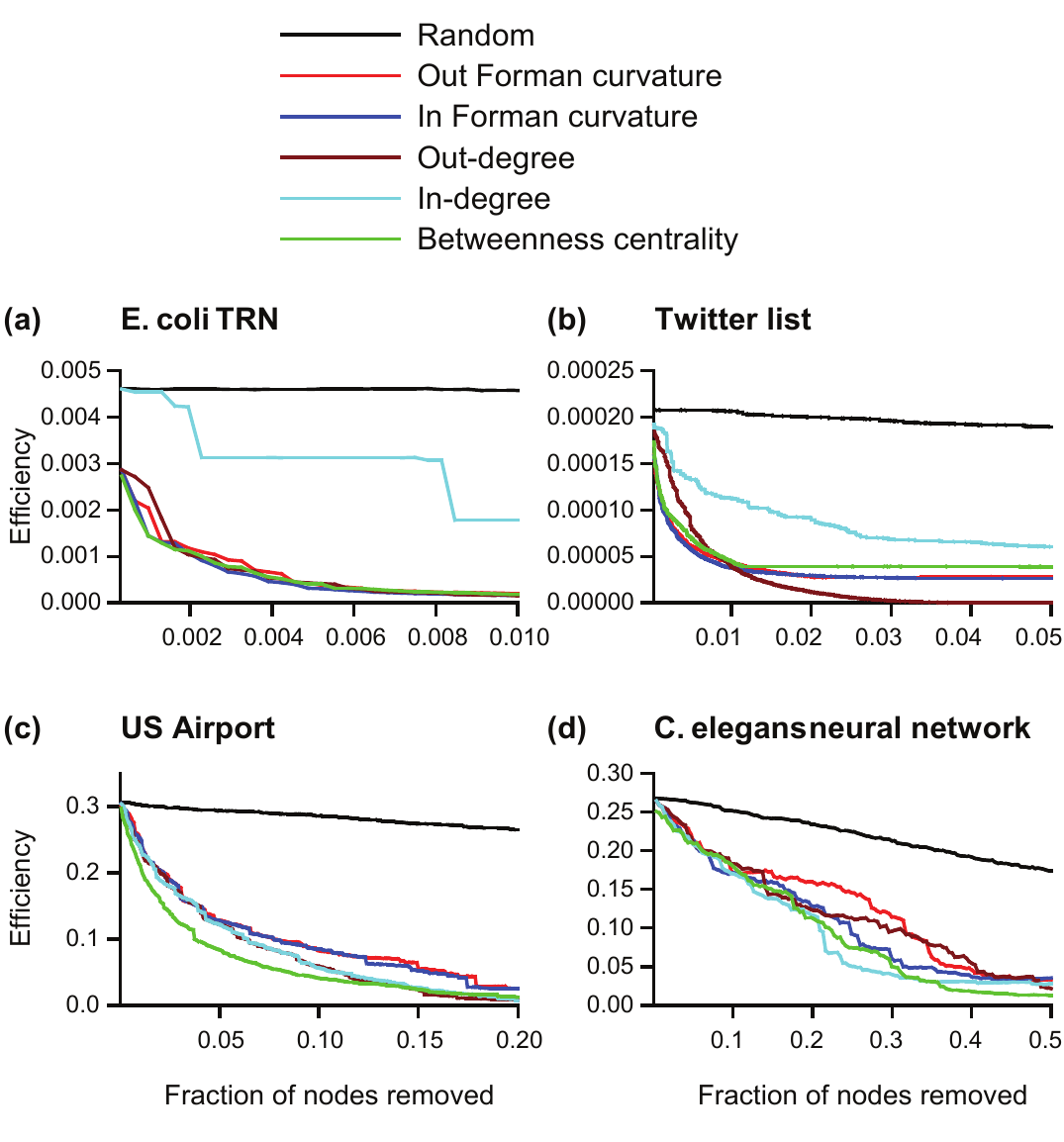}
\caption{Communication efficiency as a function of the fraction of nodes removed in real directed networks. (a) \textit{E. coli} TRN. (b) Twitter Lists. (c) US Airport. (d) \textit{C. elegans} neural network.  In this figure, the order in which the nodes are removed is based on the following criteria: Random order, Increasing order of In Forman curvature, Increasing order of Out Forman curvature, Decreasing order of in-degree, Decreasing order of out-degree, and Decreasing order of betweenness centrality.}
\end{figure}
%-----------------------------------------------------------------

%-----------------------------------------------------------------
\end{document}